\documentclass[aps,prl,twocolumn,groupedaddress,superscriptaddress,showpacs]{revtex4}
\usepackage{graphicx}
\begin{document}

\def\et{$\it{et~al.}$}
\def\He{$^{5}_{\Lambda}$He}
\def\Li{$^{6}_{\Lambda}$Li}
\def\a{$\alpha_{p}^{NM}$}
\def\am{$\alpha^{M}$}
\def\L{$\Lambda$}
\def\GeVc{GeV/$c$}
\def\P{$P_{\Lambda}$}
\def\pip{($\pi^{+}$, $p$)}
\def\GnGp{$\Gamma_{n}/\Gamma_{p}$}
\def\Eth{$E^{th}_{p}$}
\def\pik{($\pi^+$, $K^+$)}


\title{Exclusive proton asymmetry measurement in non-mesonic weak decay of polarized $^{5}_{\Lambda}$He.}


\author{T. Maruta}
\affiliation{%
  Department of Physics, University of Tokyo, Tokyo 113-0033, Japan}
\author{S. Ajimura}
\affiliation{%
  Department of Physics, Osaka University, Osaka 560-0043, Japan}
\author{K. Aoki}
\affiliation{%
  High Energy Accelerator Research Organization (KEK), Tsukuba, Ibaraki 305-0801, Japan}
\author{A. Banu}
\affiliation{%
  GSI, Darmstadt D-64291, Germany}
\author{H. Bhang}
\affiliation{%
  Department of Physics, Seoul National University, Seoul 151-742, Korea}
\author{T. Fukuda}
\altaffiliation{%
  Present address: Osaka Electro-Communication University, Osaka 572-8530, Japan}
\affiliation{%
  High Energy Accelerator Research Organization (KEK), Tsukuba, Ibaraki 305-0801, Japan}
\author{O. Hashimoto}
\affiliation{%
  Physics Department, Tohoku University, Sendai 980-8578, Japan}
\author{J. I. Hwang}
\affiliation{%
  Department of Physics, Seoul National University, Seoul 151-742, Korea}
\author{S. Kameoka}
\affiliation{%
  Physics Department, Tohoku University, Sendai 980-8578, Japan}
\author{B. H. Kang}
\affiliation{%
  Department of Physics, Seoul National University, Seoul 151-742, Korea}
\author{E. H. Kim}
\affiliation{%
  Department of Physics, Seoul National University, Seoul 151-742, Korea}
\author{J. H. Kim}
\altaffiliation{%
  Present address: Korea Research Institute of Standards and Science (KRISS), Daejeon, 305-600, Korea}
\affiliation{%
  Department of Physics, Seoul National University, Seoul 151-742, Korea}
\author{M. J. Kim}
\affiliation{%
  Department of Physics, Seoul National University, Seoul 151-742, Korea}
\author{Y. Miura}
\affiliation{%
  Physics Department, Tohoku University, Sendai 980-8578, Japan}
\author{Y. Miyake}
\affiliation{%
  Department of Physics, Osaka University, Osaka 560-0043, Japan}
\altaffiliation{%
  Present address: Department of Physics, University of Tokyo, Tokyo 113-0033, Japan}
\author{T. Nagae}
\affiliation{%
  High Energy Accelerator Research Organization (KEK), Tsukuba, Ibaraki 305-0801, Japan}
\author{M. Nakamura}
\affiliation{%
  Department of Physics, University of Tokyo, Tokyo 113-0033, Japan}
\author{S. N. Nakamura}
\affiliation{%
  Physics Department, Tohoku University, Sendai 980-8578, Japan}
\author{H. Noumi}
\affiliation{%
  High Energy Accelerator Research Organization (KEK), Tsukuba, Ibaraki 305-0801, Japan}
\author{S. Okada}
\altaffiliation{%
  Present address: RIKEN, Wako, Saitama 351-0198, Japan}
\affiliation{%
  Department of Physics, Tokyo Institute of Technology, Tokyo 152-8551, Japan}
\author{Y. Okayasu}
\affiliation{%
  Physics Department, Tohoku University, Sendai 980-8578, Japan}
\author{H. Outa}
\altaffiliation{%
  Present address: RIKEN, Wako, Saitama 351-0198, Japan}
\affiliation{%
  High Energy Accelerator Research Organization (KEK), Tsukuba, Ibaraki 305-0801, Japan}
\author{H. Park}
\affiliation{%
  Korea Research Institute of Standards and Science (KRISS), Daejeon, 305-600, Korea}
\author{P. K. Saha}
\altaffiliation{%
  Present address: Japan Atomic Energy Research Institute, Tokai, Ibaraki 319-1195, Japan}
\affiliation{%
  High Energy Accelerator Research Organization (KEK), Tsukuba, Ibaraki 305-0801, Japan}
\author{Y. Sato}
\affiliation{%
  High Energy Accelerator Research Organization (KEK), Tsukuba, Ibaraki 305-0801, Japan}
\author{M. Sekimoto}
\affiliation{%
  High Energy Accelerator Research Organization (KEK), Tsukuba, Ibaraki 305-0801, Japan}
\author{T. Takahashi}
\altaffiliation{%
  Present address: High Energy Accelerator Research Organization (KEK), Tsukuba, Ibaraki 305-0801, Japan}
\affiliation{%
  Physics Department, Tohoku University, Sendai 980-8578, Japan}
\author{H. Tamura}
\affiliation{%
  Physics Department, Tohoku University, Sendai 980-8578, Japan}
\author{K. Tanida}
\altaffiliation{%
  Present address: RIKEN, Wako, Saitama 351-0198, Japan}
\affiliation{%
  Department of Physics, University of Tokyo, Tokyo 113-0033, Japan}
\author{A. Toyoda}
\affiliation{%
  High Energy Accelerator Research Organization (KEK), Tsukuba, Ibaraki 305-0801, Japan}
\author{K. Tsukuda}
\affiliation{%
  Physics Department, Tohoku University, Sendai 980-8578, Japan}
\author{T. Watanabe}
\affiliation{%
  Physics Department, Tohoku University, Sendai 980-8578, Japan}
\author{H. J. Yim}
\affiliation{%
  Department of Physics, Seoul National University, Seoul 151-742, Korea}

\date{\today}

\begin{abstract}
 The asymmetry parameter {\a} for a proton exclusively 
 emitted in the $\Lambda p \rightarrow np$
 process was, for the first time, measured in the non-mesonic weak decay of a polarized {\He}
hypernucleus by selecting the proton-neutron
 pairs emitted in the back-to-back kinematics.
 The highly polarized {\He} was abundantly produced with the {\pik} reaction
 at 1.05 {\GeVc} in the scattering angular range of $\pm$15$^{\circ}$.
 The obtained value {\a}=0.31$\pm$0.22, as well as that for inclusive 
 protons, {\a}=0.11$\pm$0.08$\pm$0.04, largely contradicts recent theoretical
 values of around $-0.6$, although these calculations well reproduce the
 branching ratios of non-mesonic weak decay. 

\end{abstract}


\maketitle

In free space, a {\L} hyperon mainly decays through a mesonic
decay process ({\L} $\rightarrow N\pi$), in which the momentum
transfer ($\Delta q$) is about 100 MeV/$c$. 
On the other hand, the non-mesonic weak decay process 
(NMWD : $\Lambda N$ $\rightarrow$ $NN$) 
becomes dominant because of the large momentum transfer ($\Delta q\sim$ 400 MeV/$c$).
Since it is experimentally very difficult to carry out two-body scattering measurements
in $\Lambda N\rightarrow NN$ or $NN\rightarrow\Lambda N$ reactions
at this moment,
studying the NMWDs of {\L} hypernuclei gives us a very unique opportunity
to investigate a new type of baryon-baryon weak interaction.
It also has an advantage that both parity conserving and parity non-conserving
parts can be studied through the flavor-changing weak process; the former part is
usually masked by the strong interaction in the case of $NN$ weak processes.

There have been two experimental observables used to investigate the NMWD process.
One is the ratio of the partial decay width between neutron-induced decay,
$\Gamma_{n}$ ($\Lambda n$ $\rightarrow$ $nn$),
and that of proton-induced decay,
$\Gamma_{p}$ ($\Lambda p$ $\rightarrow$ $np$), called as the {\GnGp} ratio.
It is sensitive to the isospin structure of the $\Lambda N$ weak interaction,
because it allows only the final isospin, $I_{f}$ = 1, in neutron-induced decay,
while $I_{f}$ = 0 and 1 are allowed in proton-induced decay.
The other is the asymmetry parameter of the decay
proton from the NMWD process ($\alpha^{NM}_{p}$).
It is obtained from
the angular distribution $W(\theta)$ of decay particles emitted from
polarized hypernuclei, as follows:
\begin{eqnarray}
W(\theta) \propto 1 + A\cos\theta = 1 + \alpha P_{\Lambda}\cos\theta ,
\label{eq:one}
\end{eqnarray}
where $A$ is the asymmetry, {\P} denotes the polarization of 
a {\L} hypernucleus and $\theta$ is the emission angle of decay particles
with respect to the polarization axis. 

The asymmetry parameter comes from the interference
between different final parity and isospin states.
For example, concerning the $s$-shell hypernucleus {\He}, 
the decay amplitudes are classified to six amplitudes according to the final spin states
of the $NN$ system, as listed
 in Table~\ref{tab:decay_amp} [\onlinecite{six_amplitude}].
Three amplitudes ($a$, $b$, $f$) have the final-state isospin $I_{f}$ = 1,
and the others ($c$, $d$, $e$) have  $I_{f}$ = 0.
Then, the asymmetry parameter 
$\alpha_p^{NM}$ is expressed as [\onlinecite{asym_eq}]
\begin{eqnarray}
\alpha_p^{NM} = \frac
      {2\sqrt{3}Re[ae^*-\frac{1}{\sqrt{3}}b(c-\sqrt{2}d)^{*}+f(\sqrt{2}c+d)^{*}]}
      {|a|^2+|b|^2+3(|c|^2+|d|^2+|e|^2+|f|^2)}.
\label{eq:two}
\end{eqnarray}


\begin{table}
  \caption{\label{tab:decay_amp}Six channels of NMWD start from the $\Lambda N$
  relative S-states.}
  \begin{ruledtabular}
    \begin{tabular}{ccccc}
      $\Lambda N$ & $NN$ & amplitude & $I_{f}$ & parity conservation\\
      \hline
      $^{1}S_{0}$ & $^{1}S_{0}$ & $a^2$ & 1 & yes \\
                  & $^{3}P_{0}$ & $b^2$ & 1 & no \\
      $^{3}S_{1}$ & $^{3}S_{1}$ & $c^2$ & 0 & yes \\
                  & $^{3}D_{1}$ & $d^2$ & 0 & yes \\
                  & $^{1}P_{1}$ & $e^2$ & 0 & no \\
                  & $^{3}P_{1}$ & $f^2$ & 1 & no \\
    \end{tabular}
  \end{ruledtabular}
\end{table}

So far, much attention has been paid to the former one {\GnGp} both
experimentally and theoretically, because early measurements
suggested {\GnGp} values $\agt$ 1, while naive theoretical estimates
based on a one-pion-exchange model gave small values of around 0.1 [\onlinecite{alberico_review}].
This large discrepancy seems to be resolved with a new measurement [\onlinecite{kang_gngp}]
 and new theoretical calculations that take account of kaon exchange etc.
[\onlinecite{parreno_calc,sasaki_calc}]; 
 both now agree with the {\GnGp} to be $0.4 \sim 0.5$.
However, it has been realized that such calculations happen to predict
large negative values of around $-0.6\sim-0.7$ for the decay asymmetry parameter \a.
This significantly contradicts with a measured \a value of 0.24$\pm$0.22 for {\He} by 
Ajimura {\et} in the KEK E278 experiment [\onlinecite{ajimura_asym}].

There could be several effects to reduce the amplitude of the measured value.
One of the effects is the final-state interaction (FSI) for nucleons
emitted in the $\Lambda N\rightarrow NN$ processes.
The detected proton could not be a proton directly emitted from
the $\Lambda p\rightarrow np$ process, but secondary
protons that underwent by FSIs, for which the initial asymmetry information would
be lost.
Another effect is the possible contamination of a two-nucleon induced
decay process ($\Lambda pN \rightarrow npN$) in which 
the proton decay asymmetry would be very small
because emitted nucleons are assumed to be 
distributed in three-body phase space uniformly.
These protons from the secondary processes and two-nucleon induced
process should have lower energies compared with the proton energy
in the direct process.
A previous measurement set a rather high detection threshold at
$\sim$ 60 MeV for protons to avoid possible contaminations of 
those low-energy protons. However, the ambiguity on the 
contaminations left a substantial systematic error.

In the present experiment KEK E462, we aimed at a high-statistics 
measurement of {\a} for {\He} together with a high detection efficiency
for neutrons.
It enabled us not only to measure the {\a} for protons directly coming from
the $\Lambda p\rightarrow np$ process by requiring a neutron emitted
in the opposite direction, but also to investigate the proton energy-threshold
dependence of the {\a} values.



The experiment was performed 
at the K6 beam line of the KEK 12-GeV proton synchrotron (PS) 
with the high-resolution and large-acceptance
superconducting kaon spectrometer (SKS) [\onlinecite{fukuda_sks}]. 
A polarized {\L} hypernucleus {\He} was produced in the $^6$Li($\pi^+$, $K^+$)
reaction at 1.05 {\GeVc} [\onlinecite{bando_pol}] 
in the scattering angle range of $\pm$15$^{\circ}$ in the horizontal
reaction plane.
In the reaction, the ground state of {\Li}, which is proton unbound,
 is initially produced, and soon decays into {\He} by emitting a proton.
Therefore, the production of the {\He} was assured by tagging the
production of the ground state of {\Li} in the {\pik} missing-mass
spectrum.

The polarization of the {\He} hypernucleus was measured from
the decay asymmetry of $\pi^-$s $A_{\pi}$ emitted from the mesonic weak decay
of {\He} by using two decay counter systems symmetrically installed
above and below the Li target.  
Since the {\He} is composed of an $\alpha$ cluster and a {\L} hyperon,
it would be a good approximation to assume that the asymmetry parameter
of the mesonic weak decay $\alpha_{\pi}^{M}$ is almost the same as that in
free space [\onlinecite{motoba}], i.e. {\am} = $-0.642\pm0.013$.
Then, the polarization of the {\He} is obtained from the measured
decay asymmetry $A_{\pi}$.

The proton decay asymmetry $A_p$ was measured at the same time by 
using the same decay counter systems.

Each decay counter system consisted of a set of drift chambers,
two sets of timing counters and a neutron counter array
with six layers of plastic counters.
Both charged and neutral decay particles were identified by them.
Concerning charged particles, pions, protons and deuterons, 
which were observed for the first time, were clearly separated.
Neutrons were well-separated from gamma rays and we selected neutrons 
with the energy from 5 MeV to 150 MeV.
The details of the experimental setup and particle identification are 
described in Ref.~\onlinecite{e462_setup}.

By selecting the ground-state region from $-4$ to $4$ MeV in excitation energy,
we succeeded to detect 5.2$\times$10$^4$ events of {\He} in the ground state.
The statistics are several-times higher than that of the previous experiment.


In fact, the asymmetry $A$ in Eq.~\ref{eq:one} was obtained 
from the up/down ratio between the yield in the up counter system $N_{U}$
and that in the down counter system $N_{D}$.
Owing to the large acceptance of the SKS spectrometer, 
we can polarize the hypernucleus upward (downward) by selecting
the {\pik} scattering direction on the left (right) in one setup.
Thus, we can take the double ratio $R$ as
\begin{eqnarray}
   R=\left(\frac{N^{+}_{U}\times{N^{-}_{D}}}
  {N^{-}_{U}\times{N^{+}_{D}}}\right)^{\frac{1}{2}}=\frac{1+A\cos(0+\varepsilon)}{1+A\cos({\pi}-\varepsilon)}=\frac{1+A\eta}{1-A\eta},
\label{eq:three}
\end{eqnarray}
where $N^{+(-)}_{U(D)}$ presents the yield in the up (down) 
counter system at the scattering angle to the left (right).
Since the decay counter systems have some finite solid angles, the average measured angle $\langle \theta \rangle$ can not be 0 and $\pi$,
and is shifted by some fraction $\varepsilon$.
In this expression, systematic errors coming from the 
difference of the detection efficiencies and the acceptances 
between the up and down decay counter systems are canceled out.
This first-order cancellation was examined 
in an analysis with the ($\pi^+$, $pX$) reaction, 
where $X$ is either a proton or a pion, simultaneously obtained with the
{\pik} reaction.
No asymmetry is expected for protons and pions in this reaction
and we confirmed it within an error of less than 0.3\% for all
the kaon scattering angles, as listed in Table~\ref{tab:result_summary}.

At first, the polarization of {\He} {\P} was obtained as
$ P_{\Lambda} = A_{\pi} / \alpha_{\pi}^{M}$ from the mesonic decay process
of the ground state,
in which ${\alpha_{\pi}^{M}}$ is assumed to be almost the same as that in free space.
The obtained polarizations {\P}'s are listed in Table~\ref{tab:result_summary}.
They are larger than  
the previous result [\onlinecite{ajimura_pol}],
particularly in the large scattering angle region.

\begin{figure}
  \includegraphics[width=7.0cm]{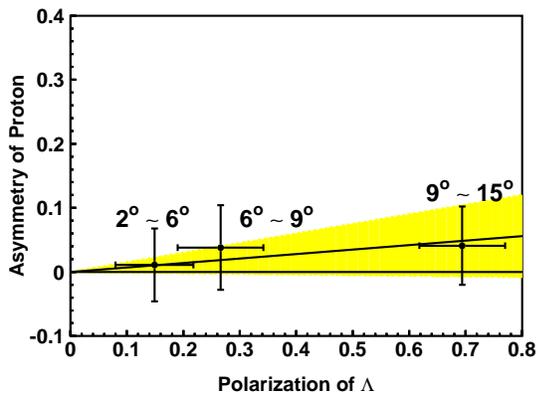}%
  \caption{\label{fig:asympara_e462}Result of proton asymmetries 
    for different scattering angles. 
    Since the horizontal axis denotes the polarization of {\He} 
    and the vertical one is the proton asymmetry,
    {\a} is obtained from a linear fit of two scattering angle regions,
    6$^{\circ}< |\theta_{K}|<$9$^{\circ}$ and
    9$^{\circ}< |\theta_{K}|<$15$^{\circ}$
  }
\end{figure}

Then, using these {\P} values, {\a} was obtained from 
the asymmetry of proton $A_{p}$ from the NMWD process of the {\He} ground state
as $\alpha_{p}^{NM} = A_{p} / P_{\Lambda}$.
Figure~\ref{fig:asympara_e462} shows the result of the asymmetries
of decay proton as a function of the polarization {\P}
measured at three different kaon scattering angles in the Lab. system 
($\theta_{K}$): 2$^{\circ}< |\theta_{K}|<$6$^{\circ}$,
6$^{\circ}< |\theta_{K}|<$9$^{\circ}$ and 
9$^{\circ}< |\theta_{K}|<$15$^{\circ}$.
From a linear fitting, the {\a} value was obtained 
to be 0.07$\pm$0.08 for 6$^{\circ}< |\theta_{K}|<$15$^{\circ}$.
The systematic error was estimated by assuming the maximum pion contamination
into protons. Since the pion asymmetry parameter has a large negative value 
 ({\am} = $-0.642\pm0.013$), the contamination tends to reduce {\a}.
Even by taking account of this effect, we obtained {\a} to be
0.11$\pm$0.07$\pm$0.04, and confirmed that {\a} is
slightly positive and small with better statistical accuracy.

\begin{figure}
 \includegraphics[width=7.0cm]{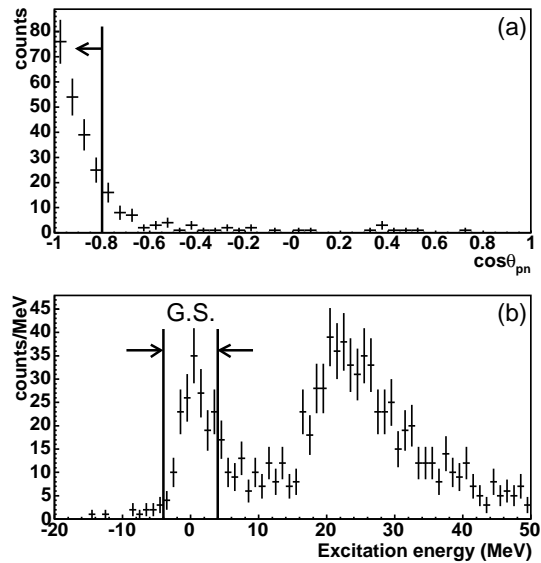}
  \caption{(a) Opening-angle distribution of a neutron and a proton from the
  non-mesonic decay of {\He}. Back-to-back events are clearly seen. 
  (b) An excitation energy spectrum with the proton-neutron pair emitted in
  the back-to-back direction.\label{fig:np}}
\end{figure}

Up to this point, all of the protons detected in the decay counter
systems, of which detection energy threshold was about 30 MeV, 
 were included in the analysis.
As already discussed, there could be some protons not directly
emitted from the $\Lambda p\rightarrow np$ process, which
might reduce the observed asymmetry of protons.
In Fig.~\ref{fig:np} (a), the opening angle distribution between
a proton and a neutron emitted from the {\He} in coincidence
is shown. The peak structure at $\cos \theta_{np} \sim -1$
is attributed to the proton-neutron pairs directly emitted from
the $\Lambda p\rightarrow np$ process, as expected from
two-body kinematics.
An excitation energy spectrum with proton-neutron pairs
in the back-to-back direction ($\cos \theta_{np} < -0.8$) is also 
shown in Fig.~\ref{fig:np} (b). The ground-state peak was
clearly identified with very small background.
The proton decay asymmetry parameter {\a} was obtained with
these back-to-back coincidence events with the same procedure
as for the inclusive proton events, and found to be 0.31$\pm$0.22.
This is the first derivation of the {\a} value for the exclusive process
$\Lambda p\rightarrow np$.

The threshold-energy dependence of the {\a} values was further
scrutinized by changing the proton detection energy thresholds as
40, 60, and 80 MeV (see Table 2). When protons are coming from the
$\Lambda p\rightarrow np$ process, the proton energy distribution
should have a broad bump structure peaked at $\sim$ 75 MeV.
The effects of FSIs and two-nucleon induced decay mode 
($\Lambda pN \rightarrow npN$) would contribute mainly
at low energy and reduce the proton decay asymmetry.
In a recent calculation by Alberico et al. [\onlinecite{alberico_calc}],
it was demonstrated that the {\a} central values were reduced by 16\%
with the 30 MeV threshold, as compared to that with the 70 MeV
threshold, due to the FSI effect introduced in their calculation.
On the contrary, we obtained rather stable {\a} values

\begin{table}
  \caption{\label{tab:result_summary}Summary table.}
  \begin{ruledtabular}
    \begin{tabular}{ccccc}
          &&2$^{\circ}\sim$ 6$^{\circ}$&6$^{\circ}\sim$ 9$^{\circ}$&9$^{\circ}\sim$ 15$^{\circ}$\\\hline
      {\pip}&$A_{p}$&$-0.000\pm$0.002&0.003$\pm$0.002&0.003$\pm$0.002\\
          &$A_{\pi}$&$-0.001\pm$0.001&0.003$\pm$0.001&0.000$\pm$0.001\\\hline
      {\pik}&{\P}&0.149$\pm$0.069&0.266$\pm$0.076&0.694$\pm$0.076\\
          &{\a}&&\multicolumn{2}{c}{0.11$\pm$0.08$\pm$0.04}\\
      $E^{th}_{p}$=40 MeV&{\a}&&\multicolumn{2}{c}{0.12$\pm$0.09$\pm$0.05}\\
      $E^{th}_{p}$=60 MeV&&&\multicolumn{2}{c}{0.20$\pm$0.10$\pm$0.05}\\
      $E^{th}_{p}$=80 MeV&&&\multicolumn{2}{c}{0.09$\pm$0.14$\pm$0.04}\\
      $np$ coincidence&{\a}&&\multicolumn{2}{c}{0.31$\pm$0.22}\\
    \end{tabular}
  \end{ruledtabular}
\end{table}

As for theoretical models based on the one-meson-exchange mechanism,
the introduction of one-kaon-exchange increased the contribution of
the amplitude $f$ in Table~\ref{tab:decay_amp}, and solved the
{\GnGp} puzzle. However, at the same time it increased the contribution of the
third term in Eq.~\ref{eq:two}, and came up with a large negative value in 
{\a} of $-0.6 \sim -0.7$. In this framework, it seems to be very hard to resolve the situation.
A recent calculation of K.~Sasaki {\et} [\onlinecite{sasaki2_calc}]
suggests that there could be a solution by further introducing the
one-sigma-exchange mechanism of a scaler-isoscaler meson $\sigma$.
It reduces the effect of the third term in Eq.~\ref{eq:two} without changing the
branching ratios very much.


In summary, we obtained the proton asymmetry parameter
in the non-mesonic weak decay of the polarized {\He} hypernucleus
in the exclusive process $\Lambda p\rightarrow np$.
The process was selected by observing the proton-neutron pairs
emitted in the back-to-back kinematics in coincidence.
The obtained value of {\a}=0.31$\pm$0.22, together with that
for the inclusive protons, {\a}=0.11$\pm$0.08$\pm$0.04,
largely deviates from the theoretical values of around $-0.6$, although
these theoretical models well reproduce the {\GnGp} ratio.
The stability of the experimental {\a} values with different proton energy
thresholds was also confirmed.
This seems to suggest the importance of a scaler-type meson-exchange
mechanism, such as a scaler-isoscaler $\sigma$ meson or a correlated
$\pi$-$\pi$ pair.


We would like to thank  Prof. K. Nakamura and the KEK-PS staff for
their support to the present experiment. We also acknowledge the
staff members of the cryogenic group at KEK.
This work was supported in part under the Korea-Japan
collaborative research program of KOSEF(R01-2005-000-10050-0)
and KRF(2003-070-C00015).

\end{document}